\documentclass[
reprint,
amsmath,
amssymb,
aps,
pra,
]{revtex4-2} 

\usepackage{graphicx}
\usepackage{dcolumn}
\usepackage{bm}
\usepackage[hidelinks]{hyperref}

\usepackage{xcolor}
\usepackage{overpic}

\renewcommand{\d}{\mathrm{d}}
\newcommand{\ee}{\mathrm{e}}
\newcommand{\ii}{\mathrm{i}}

\newcommand{\bra}[1]{\langle{#1}|}
\newcommand{\ket}[1]{|{#1}\rangle}

\begin{document}
	\title{Improving Transmon Qudit Measurement on IBM Quantum Hardware}
	
	\author{Tobias Kehrer}\email{tobias.kehrer@unibas.ch}
	\author{Tobias Nadolny}
	\author{Christoph Bruder}
	\affiliation{Department of Physics, University of Basel, Klingelbergstrasse 82, CH-4056 Basel, Switzerland}
	
	\date{\today}
	
	\begin{abstract}
		The Hilbert space of a physical qubit typically features more than two energy levels. Using states outside the qubit subspace can provide advantages in quantum computation. To benefit from these advantages, individual states of the $d$-dimensional qudit Hilbert space have to be discriminated during readout. We propose and analyze two measurement strategies that improve the distinguishability of transmon qudit states. Based on a model describing the readout of a transmon qudit coupled to a resonator, we identify the regime in hardware parameter space where each strategy is optimal. We discuss these strategies in the context of a practical implementation of the default measurement of a ququart on IBM Quantum hardware whose states are prepared by employing higher-order $X$ gates that make use of two-photon transitions.
	\end{abstract}
	
	\maketitle

	\noindent

	\section{Introduction}
	Conventional quantum computing is based on qubits which are realized on two-level subspaces of a larger physical Hilbert space. A number of physical realizations of qubits have been proposed and implemented on various platforms. These include superconducting qubits~\cite{cQED}, trapped ions~\cite{trapped_ion_review}, cold atoms and Rydberg atoms~\cite{rydberg_atoms_review}, as well as electron spins in quantum dots~\cite{spin_qubits_review}. On all of these platforms, it is necessary to isolate the two-dimensional qubit space from the remaining states of the physical Hilbert space to avoid leakage out of the computation space. However, utilizing qudits, i.e., $d$-dimensional building blocks of quantum computation, can provide advantages \cite{QuditComputingReview, AncillaFree, QuditGatesEgger2022, QuditMolecules, Shelving1, Toffoli, GHZqutrit, WalshHgate, FastReset1, FastReset2, QuditSU3, QutritBOPM, QutritNHQC}. Examples include implementing an ancilla qubit within the second and third excited states of a qudit \cite{AncillaFree}. Another example is the so-called shelving \cite{Shelving1}: by transferring the population of the first excited state to the second excited state, the error of identifying the ground state decreases.
	
	Superconducting qubits \cite{Wallraff2004, Koch2007} are prominent building blocks of noisy intermediate-scale quantum systems. The most promising example is the so-called transmon that can effectively be described as a quantum anharmonic electromagnetic oscillator. In this system, the two lowest-energy levels are identified as the qubit. Taking into account higher-lying transmon levels leads to a natural realization of a superconducting qudit. The smallest extension of the qubit is the qutrit, i.e., a three-level system. Qutrits have been used to implement a Toffoli gate \cite{Toffoli} with a significantly lower number of elementary gates compared with a realization based on two-level systems. Another interesting example is the recent experimental demonstration of a qutrit Greenberger–Horne–Zeilinger (GHZ) state \cite{GHZqutrit}.
	
	In general, if one is interested in measuring a qudit state, a proper classification of all levels involved is needed. In \cite{WallraffTimeTrace2010}, the qubit state is determined by a fit of the time evolution of the system. In setups which do not provide time-resolved data, such as the current IBM Quantum \cite{IBMQ} devices, other methods of separating qudit states have to be employed \cite{TwoToneReadout, QuditPositions, Ququart}. The strategies described in both \cite{TwoToneReadout} and \cite{QuditPositions} involve exciting the qudit-resonator system at readout drive frequencies other than the default frequency. At the default frequency, the distinguishability of the ground state and first excited state is maximized, whereas using the adapted frequencies aims at optimizing distances between different pairs of qudit states.

	In this paper, we propose and evaluate improvements of the measurement scheme of transmon qudit states by enhancing their distinguishability. To optimize the readout, we determine the measurement errors from the assignment matrix whose entries denote the probability to classify a measurement outcome to a state $\ket{i}$ even if state $\ket{j}$ was prepared. This assignment matrix is calculated using qudit-state dependent resonator steady-state amplitudes obtained from a model describing the readout of a transmon qudit by driving a coupled resonator. The default measurement schedule of most superconducting quantum hardware consists of a single-tone drive applied to the readout resonator. The frequency of the tone is chosen to maximally separate the ground and first excited states. The strategies we propose are based on modified readout resonator drive frequencies that take into account the separation of all qudit states. These strategies include a single-frequency as well as a multifrequency readout scheme. For a ququart, viz., the four lowest states of a qudit, we compare the proposed strategies in simulation and show that depending on hardware parameters, both strategies can be beneficial. We furthermore compare the model to a measurement of the drive-frequency-dependent resonator states on a current IBM Quantum device.
	
	The paper is organized as follows. In Section~\ref{sec:model}, we present a mean-field model describing the readout sequence of a transmon qudit coupled to a harmonic readout resonator. Based on this model, we calculate the readout drive-frequency-dependent assignment errors between multiple states that in some limits can be expressed analytically. In Section~\ref{sec:strategies}, we analyze both proposed readout schemes that aim to minimize these errors. We compare the data that we generated on current IBM Quantum hardware, see Section~\ref{sec:measurement}, to the readout model and strategies discussed in Sections~\ref{sec:model} and \ref{sec:strategies}. To improve the state preparation required in this procedure, we propose to add two-photon transitions to the universal gate set of qudit gates \cite{QuditGatesEgger2022} and show that this will reduce the execution time of certain qudit circuits and the duration of $X$-gate calibrations, see Appendices~\ref{sec:QuditDrivePerturbation} and \ref{sec:StatePreparation}. Finally, we conclude in Section~\ref{sec:conclusion}.

	\section{Transmon-Resonator System}\label{sec:model}
	\subsection{Effective Hamiltonian}
	The fundamental building blocks of a superconducting quantum computer are a quantum anharmonic oscillator, i.e., the transmon qudit, coupled to a harmonic oscillator, i.e., the readout resonator. Following \cite{Koch2007}, the corresponding effective Hamiltonian obtained by treating the Jaynes-Cummings interaction between the qudit and the resonator as a small perturbation, see Appendix~\ref{sec:JCPerturbation}, reads
	\begin{align}
	H_\text{eff}&= \sum_j(\omega_j+\chi_{j-1, j}+\chi_j a^\dagger a)\ket{j}\bra{j}+\omega_r a^\dagger a\,.\label{eq:Heff}
	\end{align}
	The parameter $\omega_j$ is the energy (see Appendix~\ref{sec:energylevels}) of the bare qudit state $\ket{j}$, $\omega_r$ is the energy of the readout resonator, and $a^{(\dagger)}$ is its annihilation (creation) operator. Here and in the rest of the paper, we set $\hbar=1$. The second-order corrections to the qudit and resonator energies, $\chi_{j-1,j}$ and $\chi_j$, are defined in Eqs.~\eqref{eq:chijj1} and \eqref{eq:chij}. Additionally, we describe a coherent driving of the resonator at frequency $\omega_d$ by \cite{cQED} 
	\begin{align}
	H_d&=\frac{\Omega}{2}\left(\ee^{\ii\omega_dt-\ii\phi}a+\ee^{-\ii\omega_dt +\ii\phi}a^\dagger\right)\label{eq:Hd}\,,
	\end{align}
	which enables the readout of qudit states.

	\subsection{Readout of States}
	The readout of a transmon qudit, in short, consists of driving the readout resonator while recording the response signal. We model the time evolution of a general quantum state $\rho$ comprised of a qudit and its readout resonator by the following Lindblad master equation:
	\begin{align}
	\frac{\d}{\d t}\rho&=-\ii[H_\text{eff}+H_d,\rho]+\kappa\mathcal{D}[a](\rho)\,,\label{eq:LindbladEQ1}\\
	\mathcal{D}[a](\rho)&=a\rho a^\dagger-\frac12(a^\dagger a \rho+\rho a^\dagger a)\,,
	\end{align}
	where $\kappa$ is the decay rate of the resonator. Using the effective Hamiltonian given by Eq.~\eqref{eq:Heff} and assuming the qudit to be in state $\ket{j}$, we arrive at the equation of motion of the mean-field amplitude $A\equiv\langle a\rangle=\text{Tr}[a\rho]$
	\begin{align}
	\frac{\d}{\d t}A&=-\ii(\omega_r + \chi_j)A-\ii\frac{\Omega}{2}\ee^{-\ii\omega_dt+\ii\phi}-\frac{\kappa}{2}A\label{eq:MFEQ1}\,.
	\end{align}
	The fact that $A$ depends on the qudit state $\ket{j}$ is used to discriminate different qudit states. If the qudit is in a mixture or superposition of states, this measurement procedure projects the qudit onto one of its Fock states \cite{cQED}.
	
	The general form of the complex value returned by an IBM Quantum device is
	\begin{align}
	\bar{A}&=\int\limits_0^T\d t\, k(t)A\,,
	\end{align}
	where $k(t)$ encodes the kernel integration instructions, see \textit{meas\_kernel} \cite{QiskitBackendSpecifications} in \textsc{Qiskit} \cite{Qiskit}, and $T$ is the total duration of the measurement. The choice $k(t)=\exp(\ii\omega_dt)$ corresponds to integrating the measurement signal in the rotating frame of the drive, see Section~\ref{sec:rotframedrive}, whereas the choice $k(t)=\exp(\ii\omega_mt)$ corresponds to a frame rotating at an arbitrary modulation frequency $\omega_m$, see Section~\ref{sec:rotframegeneral}.
	\begin{figure}
		\includegraphics[width=8.6cm]{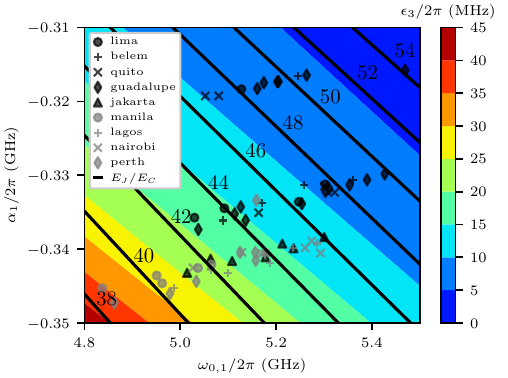}
		\caption{Overview of qudit parameters of the IBM Quantum devices listed in the legend. The qubit resonance frequency and anharmonicity are denoted by $\omega_{0,1}$ and $\alpha_1$, see Appendix~\ref{sec:energylevels}. The energy dispersion $\epsilon_3$ of the third excited state given in Eq.~\eqref{eq:epsilon} follows from these device specifications that were accessed on May 23, 2023. The labeled straight black lines denote constant values of $E_J/E_C$.}
		\label{fig:chargedisp}
	\end{figure}
	
	In this paper, we mainly consider the offset charge configuration $n_g=0$. The value of $n_g$ influences the transmon qudit energy spectrum, see Appendix~\ref{sec:energylevels}. Note that due to the significant dependence on $n_g$ of the third and higher excited states, their corresponding readout resonator states may be smeared out in phase space if charge noise is present. In Fig.~\ref{fig:chargedisp} we present an overview of the energy dispersion $\epsilon_3$ of the third excited state defined in Eq.~\eqref{eq:epsilon} for several IBM Quantum devices. Since $\epsilon_3$ decreases with increasing $E_J/E_C$, qudits that lie in the upper-right region are preferred in general.

	\subsubsection{Rotating Frame of Drive}\label{sec:rotframedrive}
	In this section, we will work in the rotating frame of the drive. Quantities in this frame will be denoted by the superscript $(d)$. Since Eq.~\eqref{eq:MFEQ1} is defined in the laboratory frame, we choose $k(t)=\exp(\ii\omega_dt)$ to transform the signal into the rotating frame of the drive and obtain
	\begin{align}
	\frac{\bar{A}^{(d)}}{T}&\stackrel{\kappa T\gg 1}{\longrightarrow}-\frac{\Omega}{2}\frac{\ee^{\ii\phi}}{\omega_r+\chi_j-\omega_d-\ii\kappa/2}\equiv A^{(d)}_j\,.\label{eq:Ad}
	\end{align}
	Here, $A^{(d)}_j$ is the steady-state resonator amplitude when the qudit is in state $\ket{j}$ and defines a coherent state $\ket{A^{(d)}_j}$. Its dependence on the resonator drive frequency is presented in Fig.~\ref{fig:phaseplottheory}a. Varying $\omega_d$, the steady-state amplitudes $A^{(d)}_j$ of the readout resonator move on a circle centered at $A_c=-\ii\ee^{\ii\phi}\Omega/2\kappa$ with diameter $\Omega/\kappa$. At resonance $\omega^{(j)}_{d,0}=\omega_r + \chi_j$, the states reach the maximum amplitude $2A_c$. For qudit readout, it is important that the distance $d_{i,j}=|A^{(d)}_i-A^{(d)}_j|$ between two qudit-state-dependent resonator states is large. We can identify two regimes of how the positions of the states in phase space depend on the readout drive frequency. For a large resonator decay rate $\kappa>|\chi_i-\chi_j|$, all states are close to the position of maximum amplitude within the same frequency range. In this case, $d_{i,j}$ exhibits only one maximum at
	\begin{align}
	\omega^{(i,j)}_{d,0}&=\omega_r+\frac{\chi_i+\chi_j}{2}\,,\\
	d_{i,j}\left(\omega^{(i,j)}_{d,0}\right)&= \frac{2\Omega|\chi_i-\chi_j|}{(\chi_i-\chi_j)^2+\kappa^2}\;.
	\end{align}
	In contrast, for a small resonator decay rate $\kappa<|\chi_i-\chi_j|$, the frequency ranges where the state amplitudes $A^{(d)}_j$ are close to the maximum amplitude do not match. Here, two drive frequencies $\omega_d=\omega^{(i,j)}_{d,\pm}$ maximize the distance $d_{i,j}$, 
	\begin{align}
	\omega^{(i,j)}_{d,\pm}&=\omega^{(i,j)}_{d,0}\pm\frac12\sqrt{(\chi_i-\chi_j)^2-\kappa^2}\,,\\
	d_{i,j}\left(\omega^{(i,j)}_{d,\pm}\right)&=\frac{\Omega}{\kappa}\equiv d_c\,,
	\end{align}
	where $d_c$ denotes the diameter of the circle on which the states move. Thus, at $\omega^{(i,j)}_{d,\pm}$, both states are located on opposite sides of the circle, which is the maximum separation they can obtain.
	
	If we set the drive frequency to $\omega^{(i,j)}_{d,0}$ (or $\omega^{(i,j)}_{d,\pm}$), i.e., maximizing the distance between state $\ket{i}$ and $\ket{j}$, the distance between other pairs of states is in general reduced and hence not optimal for discrimination of these states. Therefore, in Section \ref{sec:strategies}, we present two measurement strategies to mitigate this issue.

	\subsubsection{General Rotating Frame}\label{sec:rotframegeneral}
\begin{figure*}[t]
		\includegraphics[width=5.9cm]{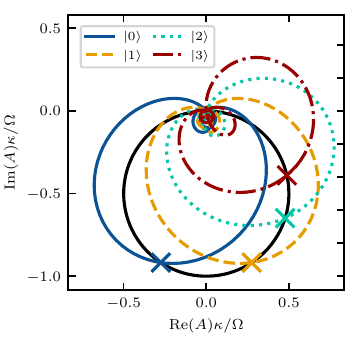}\llap{
			\parbox[b]{0.2\linewidth}{\textbf{(a)}\\\rule{0in}{1.42\linewidth}
		}}
		\includegraphics[width=5.9cm]{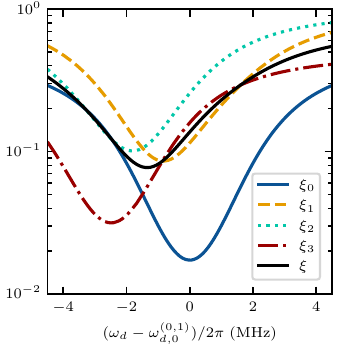}\llap{
			\parbox[b]{0.3\linewidth}{\textbf{(b)}\\\rule{0in}{0.95\linewidth}
		}}
		\includegraphics[width=5.9cm]{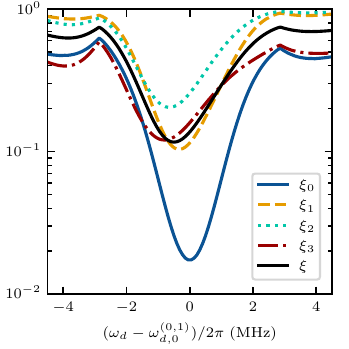}\llap{
			\parbox[b]{0.3\linewidth}{\textbf{(c)}\\\rule{0in}{0.95\linewidth}
		}}
		\caption{(a) Drive-frequency-dependent phase-space positions $A^{(d)}_j$ and $A^{(m)}_j$ of the coherent state of the resonator given the qudit prepared in $\ket{j}$, see Eqs.~\eqref{eq:Ad} and \eqref{eq:Am}. For $\omega_m=\omega_d$, the trajectories of all states $A^{(d)}_j$ match, denoted by the black circle. The colored lines denote $A^{(m)}_j$ for $\omega_m=\omega^{(0,1)}_{d,0}$. Crosses highlight the positions at $\omega_d=\omega_m$ where both models match, $A^{(d)}_j=A^{(m)}_j$. (b),(c) Error measures $\xi_j$ and $\xi$ in the frame of $\omega_d$ and $\omega_m$ respectively, see Eqs.~\eqref{eq:Ejdef} and \eqref{eq:Edef}. Following Appendices~\ref{sec:energylevels} and \ref{sec:JCPerturbation}, for these plots, we determine $E_J/E_C$ by the qubit parameters $\omega_{0,1}$ and $\alpha_1$ from \textit{ibm\_lagos Q4} (July 7, 2023). Moreover, we choose $g/2\pi=100\,$MHz, $\Omega/2\pi=100\,$MHz, $\kappa/2\pi =5\,$MHz, $T=0.35\,$\textmu s, $\sigma_j=0.13\Omega/\kappa$, $\phi=0$, and $n_g=0$.}
		\label{fig:phaseplottheory}
	\end{figure*}

	In a frame of a general rotation frequency $\omega_m$, i.e., choosing $k(t)=\exp(\ii\omega_m t)$, the state reached in the long-time limit $\kappa T\gg 1$ is time dependent,
	\begin{align}
	\frac{\bar{A}^{(m)}}{T}&\stackrel{\kappa T\gg 1}{\longrightarrow}\ee^{\ii((\omega_m-\omega_d)T/2)}\text{sinc}\left((\omega_d-\omega_m)T/2\right)
	A^{(d)}_j\nonumber\\
	&~~\equiv A^{(m)}_j\,,\label{eq:Am}
	\end{align}
	where $\text{sinc}(x)=\sin(x)/x$ and the superscript $(m)$ is used to denote quantities in this frame. The difference between Eqs.~\eqref{eq:Ad} and \eqref{eq:Am} is an additional factor of sinc peaking at $\omega_d=\omega_m$. These resonator state amplitudes and their dependence on the drive frequency $\omega_d$ are also visualized in Fig.~\ref{fig:phaseplottheory}a. The states $A^{(d)}_j$ move on the black circle with diameter $\Omega/\kappa$, whereas the motion of the states $A^{(m)}_j$ follows a distorted circle (colored lines).

	\section{Measurement Strategies}\label{sec:strategies}
	In the previous section, we presented a model describing the readout on superconducting quantum hardware. The centers of the Wigner functions of the coherent readout resonator states when the qudit is in state $\ket{j}$ are given by $A_j$. Due to intrinsic quantum noise and hardware limitations, the possible readout resonator states for each qudit state overlap. This leads to potential misclassification and thus measurement errors when reading out the qudit states.
	
	In the following, we propose two strategies for improving qudit readout compared to the default measurement scheme that utilizes a single resonator drive frequency that optimizes the classification of $\ket{0}$ and $\ket{1}$. The first strategy consists of finding a single frequency that maximizes the distinguishability between all $d$ qudit Fock states. In the second strategy, we allow for multiple different drive frequencies.

	\subsection{Assignment Matrix}
	To arrive at a measure of the distinguishability of states, we introduce the measurement assignment matrix $M$ \cite{AssignmentMatrix}. The qudit-state-dependent resonator states are defined by their steady-state amplitude $A_j$. We assume their Wigner functions to follow a two-dimensional Gaussian distribution,
	\begin{align}
	G(z, A_j, \sigma_j)&=\frac{1}{2\pi\sigma_j^2}\exp\left(-\frac{|z-A_j|^2}{2\sigma_j^2}\right)\,,\label{eq:Gaussian}
	\end{align}
	centered at $A_j$ with standard deviation $\sigma_j$ larger than the intrinsic quantum noise. The elements of $M$ are given by
	\begin{align}
	M_{i,j}&=\int\limits\d^2z\,G(z, A_j, \sigma_j) \prod_{k\neq i}\Theta_{i,k}\,
	\label{eq:AssignmentMatrix}
	\end{align}
	and define the probability to classify a measurement as state $\ket{i}$ even if state $\ket{j}$ was prepared. The region corresponding to each state $\ket{i}$ is defined by the maximum likelihood estimator (MLE) leading to
	\begin{align}
	\Theta_{i,k}&=\Theta\left(G(z, A_i, \sigma_i) - G(z, A_k, \sigma_k)\right)\,,
	\end{align}
	where $\Theta$ denotes the Heaviside function. For $\sigma_j=\sigma$ (valid assumption for this hardware setup, see the discussion about the distribution of $\sigma$ of Gaussian fits in the second paragraph of Section~\ref{sec:measurement}), the MLE is equivalent to the minimum distance estimator (MDE) that implies
	\begin{align}
	\Theta_{i,k}&=\Theta(|z-A_k| - |z-A_i|)\,.\label{eq:MDE}
	\end{align}
	Using the MDE, a data point $z$ is assigned to the region of state $A_i$ if its Euclidean distance to all of the other states $A_{k}$ is larger. In contrast, using the MLE, a data point $z$ is assigned to the region of the state $A_i$ that has the largest probability density. For simplicity and since in our measurements all $\sigma_j$ are comparable, we choose the MDE throughout this paper.
	
	Ideally, $M_{i,j}=\delta_{i,j}$, meaning perfect measurement. We define two measures $\xi_j$ and $\xi$,
	\begin{align}
	\xi_j&=1-M_{j,j}\,,\label{eq:Ejdef}\\
	\xi&=\frac{1}{d}\sum_{j=0}^{d-1}\xi_j\,,\label{eq:Edef}
	\end{align}
	where $\xi_j$ is the probability of misclassifying the qudit state $\ket{j}$. The dependence of $\xi_j$ and $\xi$ on the readout resonator drive frequency is shown in Figs.~\ref{fig:phaseplottheory}b and c. The measurement errors $\xi_j$ achieve their minima at different readout resonator drive frequencies. If $\omega_m\neq\omega_d$, the locations of the minima cannot be distinguished as well as for $\omega_m=\omega_d$. In the current setup of IBM Quantum hardware, the frequency $\omega_m$ of the rotating frame cannot be changed. Therefore, the difference between the frequency dependencies of all $\xi_j$ is less pronounced.
	
	Note that for setups where all qudit states lie on a circle (e.g., always for qutrits) and $\sigma_j=\sigma$, $M_{i,j}$ can be expressed in terms of Owen's $T$ function, see Appendix~\ref{sec:OwenT}. This allows for fast numerical calculation of Eq.~\eqref{eq:AssignmentMatrix}.

	\subsection{Finite Sampling}\label{sec:FiniteSampling}
	In experiments, measuring an unknown state $\ket{\psi}=\sum_jc_j\ket{j}$ in the $Z$ basis is equivalent to estimating its populations $p_j\equiv|c_j|^2$ based on a set of $N$ data points $\lbrace z_j\rbrace$, also called shots. For each shot, the total state is projected onto one of the $d$ qudit states $\ket{j}$ with probability $p_j$. Therefore, the total probability distribution of measuring one shot at $z$ given $\vec{p}$ is a sum of all $d$ Gaussians defined in Eq.~\eqref{eq:Gaussian} weighted by $p_j$. The measurement task can be understood as learning the parameters $\vec{p}=(p_j)$ of this multimodal probability distribution,
	\begin{align}
	P(z|\vec{p})&=\sum_{j=0}^{d-1}p_j G(z,A_j,\sigma_j)\,,\label{eq:ProbToLearn}
	\end{align}
	where $A_j$ and $\sigma_j$ are obtained from a separate measurement. The space of possible $\vec{p}$,
	\begin{align}
	p_j&\in[0,1],~~p_{d-1}=1-\sum_{j=0}^{d-2}p_j\,,
	\end{align}
	can be mapped to a $(d-1)$-simplex, using the normalization condition of $\vec{p}$.
	
	We describe the measurement analysis as follows. Each shot is labeled by a state $\ket{j}$ depending on its phase-space distance (MDE) to the $d$ qudit Fock states. The resulting list of counts $\vec{N}$ can be used to obtain information about $\vec{p}$. In Appendix D, we show that by using Bayes' formalism, the probability distribution of $\vec{p}$ given $\vec{N}$ is the Dirichlet distribution,
	\begin{align}
	P(\vec{p}\,|\vec{N})&=\frac{1}{\mathcal{N}}\text{Dir}(M\vec{p}, \vec{N})\,.\label{eq:Dir1}
	\end{align}
	The component $N_j$ of $\vec{N}$ equals the number of shots classified as state $\ket{j}$. The assignment matrix $M$ reflects the fact that some shots are classified incorrectly. Given $\vec{N}$, the location of the maximum (also called mode) can be computed analytically,
	\begin{align}
	\vec{p}_\text{mode}&=\frac{1}{N}M^{-1}\vec{N}\,.\label{eq:DirMode}
	\end{align}
	This result is similar to a common procedure known in \textsc{Qiskit} as ``measurement error mitigation''. Note that applying the inverse of $M$ to $\vec{N}$ can lead to negative components of $\vec{p}_\text{mode}$. In \textsc{Qiskit}, this problem is circumvented by approximating $\vec{p}_\text{mode}$ by the valid $\vec{p}\,'$ that is closest (in two-norm) to $\vec{N}$, see method \textit{least\_squares} in \textit{qiskit.utils.mitigation.\_filters.py} \cite{Qiskit},
	\begin{align}
	\vec{p}\,'=\underset{\vec{p}}{\text{argmin}}\left(|\vec{N}/N-M\vec{p}\,|^2\right)\,.\label{eq:pmin}
	\end{align}
	Equation~\eqref{eq:pmin} is the estimate of the state populations $\vec{p}$ after measuring $\vec{N}$ shots. We will use the uncertainty of these estimates, viz., the numerically calculated standard deviations SD$[p_j]$, to decide which of the proposed strategies performs best, i.e., exhibits the smallest standard deviation.

	\subsection{Comparison of Strategies}\label{sec:StratComp}
	We consider two strategies that make use of either one or multiple drive frequencies. In the default readout scheme of superconducting quantum hardware, measurement pulses with a single drive frequency that maximizes the distinguishability between the qubit states $\ket{0}$ and $\ket{1}$ are applied.
	
	The first strategy we propose replaces the default frequency by the one that optimally separates all qudit states in phase space simultaneously. Since, in general, the state $\ket{\psi}$ that we want to measure is unknown, we suggest to optimize $\xi$, see Eq.~\eqref{eq:Edef}, which is the average of the individual measurement errors $\xi_j$.
	
	The second strategy uses $N/d$ shots for each of the $d$ different frequencies at which individual states are most isolated, i.e., $\xi_j$ are minimal. We will show that this strategy is advantageous in cases when there is no single frequency at which all states are separated well enough. Hardware parameters and the state to be measured determine which of the two strategies outperforms the other.
	
	To compare both strategies, we draw $N=1000$ samples from the probability distribution given by Eq.~\eqref{eq:ProbToLearn} for $\sigma_j=\sigma$ and an equal-superposition state $p_j=1/d$. The drive frequencies we use are the locations of the minimum of $\xi$ for the single-drive strategy and the minima of $\xi_j$ for the multifrequency strategy. Each sample is classified using the MDE, see Eq.~\eqref{eq:MDE}, i.e., by its Euclidean distance to the nearest state $\ket{j}$. The final probability distribution for the $p_j$ of the single-frequency strategy is given in Eq.~\eqref{eq:Dir1}. The final probability distribution for the multifrequency strategy is the normalized product of the term in Eq.~\eqref{eq:Dir1} for each measurement frequency $\omega_k$,
	\begin{align}
	P(\vec{p}\,|\lbrace\vec{n}_k\rbrace)\propto\prod_{k=0}^{d-1}\mathrm{Dir}(M(\omega_k)\vec{p}\,|\vec{n}_k)\,,\label{eq:DirMultiProd}
	\end{align}
	where $\vec{n}_k$ is the list of counts of classified shots for the $k^\text{th}$ measurement frequency. The standard deviation SD$[p_j]$ is computed numerically from this distribution.

	\begin{figure}
		
		\includegraphics[width=8.6cm]{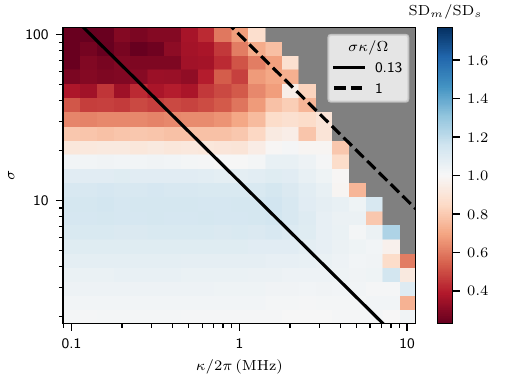}
		\caption{Ratio of the standard deviation SD$_m$ for the multifrequency strategy and the standard deviation SD$_s$ of the single-frequency strategy applied to an equal-superposition state taking $N=1000$ shots. The gray region indicates where both standard deviations exceed SD$_{s/m}\geq0.1$. The straight lines denote constant values of $\sigma\kappa/\Omega$. We take the same qudit parameters as in Fig.~\ref{fig:phaseplottheory} and choose $g/2\pi=100\,$MHz and $\Omega/2\pi=100\,$MHz.}
		\label{fig:KappaAndSigmaVsVar}
	\end{figure}

	\begin{figure*}[t]
		\begin{overpic}[width=5.9cm, tics=20]
			{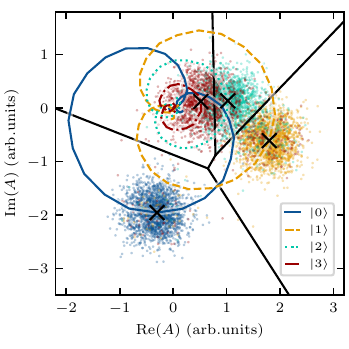}
			\put(25,89){(a)}
		\end{overpic}
		\begin{overpic}[width=5.9cm, tics=20]
			{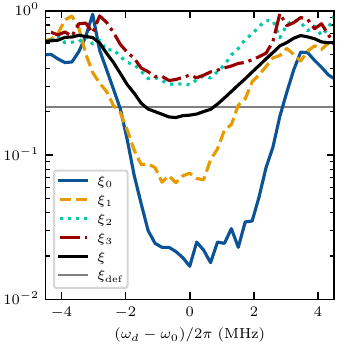}
			\put(50,89){(b)}
		\end{overpic}
		\begin{overpic}[width=5.9cm, tics=20]
			{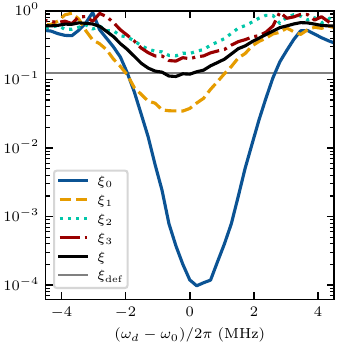}
			\put(50,89){(c)}
		\end{overpic}
		\caption{(a) Drive-frequency-dependent ququart measurements on \textit{ibm\_lagos Q4} (July 7, 2023), the experimental equivalent of Fig.~\ref{fig:phaseplottheory}a. The colored curves (with white shadows) correspond to the centers of Gaussian fits to the Fock states $\ket{j}$. Black straight lines indicate the boundaries of regions assigned to individual Fock states. These boundaries are constructed using the minimum distance estimator, see Eq.~\eqref{eq:MDE}. For the drive frequency $\omega_0=7.2463\,$GHz, we show the measurement results of all $N=2000$ shots for each prepared Fock state. Black crosses mark the centers of their Gaussian fits. (b) Measurement errors $\xi_j=1-M_{j,j}$ based on all measured shots of the data presented in (a). Here, the elements $M_{i,j}$ of the assignment matrix equal the relative number of shots, $N_i/N$, that are classified as $\ket{i}$ even if $\ket{j}$ is prepared. The horizontal gray line denotes the measurement error $\xi_\text{def}$ obtained using the default measurement pulse. (c) Measurement errors $\xi_j$ based on Gaussian fits to the data presented in (a) and the assignment matrix $M$ defined in Eq.~\eqref{eq:AssignmentMatrix}. Using the centers and average $\sigma$ of Gaussian fits for each readout resonator drive frequency, we calculate $M$ numerically. The $\xi_j$ shown in (b) are larger than those in (c) since they do not only represent assignment errors, but also include additional errors such as qudit decay, leakage, and imperfect state preparation. In both (b) and (c), the minimum of the average assignment error $\xi$ will be smaller than $\xi_\text{def}$ obtained by the default pulse if the modulation frequency is chosen such that $\omega_m=\omega_d$, cf.~Figs.~\ref{fig:phaseplottheory}b and c.}
		\label{fig:data}
	\end{figure*}

	Figure~\ref{fig:KappaAndSigmaVsVar} shows the dependence of the ratio of both averaged standard deviations,
	\begin{align}
	\mathrm{SD}_{s/m}=\frac{1}{d}\sum_{j=0}^{d-1}\mathrm{SD}_{s/m}[p_j]\,,
	\end{align}
	of $p_j$ on hardware parameters $\sigma_j=\sigma$ and $\kappa$. The blue region corresponds to setups for which the standard deviation SD$_{s}$ of $p_j$ using a single-drive frequency scheme is smaller. In contrast, the red region corresponds to hardware configurations where it is beneficial to measure at multiple frequencies, i.e., SD$_{m}$ of the multidrive frequency scheme is smaller. The gray region indicates parameter values for which both standard deviations exceed SD$_{s/m} \geq 0.1$. Since the expected values $p_j$ lie in $[0,1]$, this threshold corresponds to an uncertainty of at least 10\%. 
	
	The overall trend is that for small $\sigma$, i.e., strongly located Gaussians, the single-frequency strategy performs at a similar, slightly better level than the multifrequency strategy. The multifrequency strategy is preferable for large $\sigma$, when the overlap of the Gaussians would be too large using a single drive frequency. Intuitively, this is expected since, for small $\kappa$ and large $\sigma$, only one state is isolated from the others which group together at the origin in phase space, see discussion of regimes ${\kappa\lessgtr|\chi_j-\chi_{j+1}|}$ in Section~\ref{sec:rotframedrive}. We also added lines of constant relative uncertainty $\sigma\kappa/\Omega$. Along these lines, the Gaussian widths $\sigma$ are fixed in units of the diameter $\Omega/\kappa$ of the circle on which the states move in the rotating frame of the drive. The solid line corresponds to $\sigma=0.13\Omega/\kappa$ chosen in Figs.~\ref{fig:phaseplottheory}b and c, whereas the dashed line approximately matches the threshold of SD$_{s/m} \geq 0.1$. Following the solid black line, the standard deviation of the single-frequency strategy appears to exhibit a minimum around $\kappa/2\pi=1-2\,$MHz. For fixed $\sigma\kappa/\Omega$ and small resonator decay rates $\kappa$, the states move around the circle rather individually, whereas for large $\kappa$, the states move as a group, cf.~Section~\ref{sec:rotframedrive}.

	\section{Measurement of a Ququart}\label{sec:measurement}
	In this section, we will compare the model described in Section~\ref{sec:model} to data obtained from \textit{ibm\_lagos Q4} on July 7, 2023. We prepare the four lowest Fock states of the transmon qudit and measure them for various readout drive frequencies. In Appendix~\ref{sec:StatePreparation}, the preparation of the individual qudit Fock states is described. Sequences of $X$ gates defined as simple Gaussian pulses are used to prepare the four lowest Fock states of the transmon qudit. As mentioned earlier, we make use of higher-order $X$ gates, see Appendix~\ref{sec:QuditDrivePerturbation}. These provide a reduction of the execution time of certain quantum circuits and a reduction of the duration of $X$-gate calibrations.
	
	In Fig.~\ref{fig:data}a we show the measurements of the four lowest Fock states for various readout resonator drive frequencies. This plot is the experimental equivalent of Fig.~\ref{fig:phaseplottheory}a. For each Fock state and for each readout resonator drive frequency, we take $N=2000$ shots while keeping the other drive parameters fixed at the default values. For $\omega_0=7.2463\,$GHz, Fig.~\ref{fig:data}a shows all shots in the color of the prepared Fock state. This value of $\omega_0$ is $-5.5\,$MHz off the default frequency reported by the IBM Quantum device. Black crosses highlight the centers of the Gaussian fits. For other drive frequencies, we only plot the centers of the Gaussian fits as colored lines (with a white shadow). The straight black lines denote the boundaries of regions (defined via MDE, see Eq.~\eqref{eq:MDE}) that are assigned to one Fock state.
	
	We analyze the measurement errors in two ways. First, we define the elements $M_{i,j}$ by the relative number of shots, $N_i/N$, classified as $\ket{i}$ even if $\ket{j}$ is prepared. In this way, $M$ incorporates misclassification errors but also additional errors such as imperfect qudit state preparation. From this matrix, we obtain the errors $\xi_j$, displayed in Fig.~\ref{fig:data}b. Second, we use the centers of the Gaussian fits for each qudit state and for each value of the resonator drive frequency and a fixed value of $\sigma$ to compute the assignment matrix defined in Eq.~\eqref{eq:AssignmentMatrix}. By examining these Gaussian fits, we find a narrow distribution of the $\sigma$ values: $\sigma=(0.302\pm0.017)$ (same arbitrary units as in Fig.~\ref{fig:data}a). The resulting errors $\xi_j$ are shown in Fig.~\ref{fig:data}c. Here, the $\xi_j$ only represent errors that arise from misassignment of shots drawn from the multi-Gaussian distribution, see Eqs.~\eqref{eq:Gaussian} and \eqref{eq:AssignmentMatrix}. Since real devices feature other sources of error, e.g., qubit decay, leakage, and imperfect state preparation, the values of $\xi_j$ presented in Fig.~\ref{fig:data}b are larger than in Fig.~\ref{fig:data}c.
	
	Our model, visualized by the theory plots in Figs.~\ref{fig:phaseplottheory}a and c, shows qualitative agreement with the data presented in Figs.~\ref{fig:data}a and c. In both Figs.~\ref{fig:data}b and c, the horizontal gray line $\xi_\text{def}$ denotes the average assignment error of the four lowest Fock states using the default readout pulse and should be compared with the solid black line $\xi$. The corresponding data were taken from Rabi calibration measurements, similar to Fig.~\ref{fig:calibrations}b, at the drive amplitude that is closest to the fitted optimum.
	
	We find a dependence of the measurement errors $\xi_j$ on the readout resonator frequency as expected. The data presented in Figs.~\ref{fig:data}b and c suggest that the default measurement frequency is not ideal to separate all four qudit states. However, the minima appear at only slightly different positions. Note that the difference in positions is only small due to IBM Quantum software/hardware limitations: $\omega_m$ cannot be set to its ideal value $\omega_m=\omega_d$, see Section~\ref{sec:rotframegeneral}. We expect the impact of varying the readout resonator drive frequency to be much higher if it is possible to analyze all data in the rotating frame of the drive, compare Figs.~\ref{fig:phaseplottheory}b~and~c.
	
	In this paper, we focused on the analysis of only four qudit states since the readout of higher excited states beyond $\ket{3}$ becomes difficult for several reasons. Higher excited states are more sensitive to charge noise, see Fig.~\ref{fig:energylevels}. Since $\chi_j$ depends on the qudit spectrum $\omega_j$, charge noise leads to ambiguous steady-state amplitudes. In addition, finding a single drive frequency that properly separates all qudit states becomes more difficult with an increasing number of qudit states. For example, for the IBM Quantum device that we utilized in this paper, we estimate $\chi_1<\chi_4<\chi_2$ which indicates that the steady-state amplitude corresponding to $\ket{4}$ lies between $\ket{1}$ and $\ket{2}$. We expect that the more states are involved, the better the performance of a multifrequency strategy in comparison to a single-frequency strategy given a small $\kappa<|\chi_i-\chi_j|$, cf.~Fig.~\ref{fig:KappaAndSigmaVsVar}.

	\section{Conclusion}\label{sec:conclusion}
	We have presented a model that describes phase-space measurement data of qudit states on superconducting quantum hardware. Our model qualitatively matches the data that we generated on a current IBM Quantum device. For qudit-state preparation, we employ higher-order $X$ gates between $\ket{j}$ and $\ket{j+2}$. This scheme leads to a reduction of the execution time of qudit quantum circuits as well as of the duration of $X$-gate calibrations. Based on our model, we have compared the performance of two measurement strategies, a single-frequency and a multifrequency scheme, in simulations. For each strategy, we have identified the regime in hardware parameter space where it is optimal. The multifrequency strategy is superior when the qudit-state-dependent resonator states overlap significantly.
	
	To use the full potential of both strategies, it is necessary to adjust the modulation frequency $\omega_m$ of the device. This is currently not possible on IBM Quantum hardware. Despite these software/hardware restrictions, we still find differences in the frequency locations of the minima of the individual measurement errors $\xi_j$ and an improvement over the measurement error $\xi_\text{def}$ using the default measurement pulse. We expect a better performance of the strategies for setups that operate in the rotating frame of the drive $\omega_m=\omega_d$.
	
	In the future, adaptive measurement schemes that change the drive frequency from shot to shot or between bunches of shots may be possible. This can lead to a further improvement of transmon qudit measurements.

	\section*{Acknowledgments}
	We would like to thank D.J.~Egger and W.~Shanks for fruitful discussions about measurement and resonator spectroscopy on IBM Quantum hardware and J.~Arnold for stimulating discussions about the assignment matrix. We thank IBM for providing access to their quantum hardware infrastructure. We acknowledge financial support from the Swiss National Science Foundation individual grant (grant no. 200020 200481).

	\appendix
	\section{Transmon Energy Levels}\label{sec:energylevels}
	To estimate resonance frequencies for various transitions, we numerically compute the energy levels of the transmon qudit Hamiltonian \cite{Koch2007,cQEDDevoret},
	\begin{align}
	H_\text{T}&=4E_C\sum_n(n-n_g)^2\ket{n}\bra{n}\nonumber\\
	&~~~~-\frac{E_J}{2}\sum_n(\ket{n}\bra{n+1}+\ket{n+1}\bra{n})\,,\label{eq:HamCPB}
	\end{align}
	depending on the offset charge $n_g$ and the ratio $E_J/E_C$. The sorted eigenvalues $E_n(n_g)$ are shifted such that $E_0(0)=0$. We define the average transition frequency $\omega_{i,j}$ between $\ket{i}$ and $\ket{j}$ of both configurations $n_g=0,1/2$ as
	\begin{align}
	\omega_{i,j}&=\frac{E_{j}(0)+E_{j}(1/2)-E_{i}(0)-E_{i}(1/2)}{2(j-i)}\,,
	\end{align}
	and the frequency difference $\Delta\omega_{i,j}$ as
	\begin{align}
	\Delta\omega_{i,j}&=\frac{E_{j}(0)-E_{i}(0)-E_{j}(1/2)+E_{i}(1/2)}{j-i}\,.
	\end{align}
	The anharmonicity $\alpha_j$ and the energy dispersion $\epsilon_j$ of the transmon qudit are defined by
	\begin{align}
	\alpha_j&=\omega_{j,j+1}-\omega_{j-1,j}\,,\\
	\epsilon_j&=E_j(0)-E_j(1/2)\,.\label{eq:epsilon}
	\end{align}
	\begin{figure}[t]
		\includegraphics[width=8.6cm]{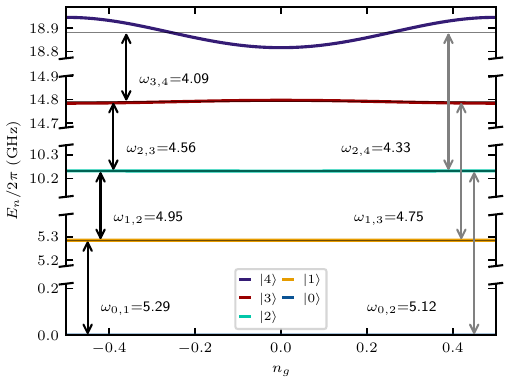}
		\caption{Numerical prediction of the energy levels $E_n$ of \textit{ibm\_lagos Q4} (July 7, 2023) based on Eq.~\eqref{eq:HamCPB} and $E_J/E_C\approx 45.6$. The transition frequencies $\omega_{i,j}$ are displayed in units of $(2\pi)\,$GHz.}
		\label{fig:energylevels}
	\end{figure}\noindent
	We numerically obtained the fundamental parameter $E_J/E_C$ of a specific IBM Quantum backend by demanding that the qubit frequency $\omega_{0,1}$ and anharmonicity $\alpha_1$ match the values reported by this device. In Fig.~\ref{fig:energylevels}, we plot the dependence of $E_n$ on $n_g$ for the five lowest states. The values of the frequency difference vary from $\Delta\omega_{0,1}/2\pi=25.1\,$kHz to $\Delta\omega_{3,4}/2\pi=-142\,$MHz. In Fig.~\ref{fig:chargedisp}, the values $\epsilon_3$ for a number of IBM Quantum devices are displayed. For large $\epsilon_j$ compared to $\omega_{j,j+1}$, the Wigner function of the state $\ket{j}$ effectively is smeared out in phase space since both configurations $n_g=0,1/2$ exhibit different resonance frequencies.

	\section{Schrieffer-Wolff Transformation}
	In this appendix, we will use Schrieffer-Wolff transformations \cite{SchriefferWolff1966} to obtain perturbative approximations of, first, the Jaynes-Cummings interaction between a qudit and a readout resonator and, second, the two-photon drive of the qudit.
	
	In general, the system of interest can be defined by the Hamiltonian
	\begin{align}
	H=H_0+\lambda H_1+\lambda V\,.
	\end{align}
	Here, $H_0$ and $H_1$ are block diagonal in the subsystems, whereas $V$ is block off diagonal. To find an effective block-diagonal Hamiltonian, i.e., eliminate the block off-diagonal part $V$, the unitary transformation $U=\ee^S$ is applied to $H$, 
	\begin{align}
	H_\text{eff}&=UHU^\dagger=\ee^{S}H\ee^{-S}\,.
	\end{align}
	Expanding the anti-Hermitian operator $S=-S^\dagger$ as $S=\sum_{n=1}^\infty\lambda^nS^{(n)}$, $H_\text{eff}$ can be expressed as
	\begin{align}
	H_\text{eff}&=H+\left[S,H\right]+\frac12\left[S,\left[S,H\right]\right]+\dots\nonumber\\
	&=\sum_{n=0}^\infty\lambda^nH^{(n)}_\text{eff}\,.
	\end{align}
	The first-order contribution reads $H^{(1)}=H_1+V+\left[S^{(1)},H_0\right]$. To eliminate the block off-diagonal $V$ in this expression, we impose $V=-\left[S^{(1)},H_0\right]$. Since $H_0$ is block diagonal, $S^{(1)}$ has to be block off diagonal. As a consequence, $\left[S^{(1)},V\right]$ is block diagonal. The second-order contribution reads $H^{(2)}=\frac{1}{2}\left[S^{(1)},V\right]+\left[S^{(1)},H_1\right]+\left[S^{(2)},H_0\right]$, and imposing $\left[S^{(1)},H_1\right]=-\left[S^{(2)},H_0\right]$ guarantees the second order to be block diagonal. The second-order contribution to the effective block-diagonal Hamiltonian is then given by
	\begin{align}
	H^{(2)}_\text{eff}&=\frac{1}{2}\left[S^{(1)},V\right]\,.
	\end{align}
	We choose a superposition of all operators appearing in $V$ as an ansatz for $S^{(1)}$.

	\subsection{Jaynes-Cummings Interaction}\label{sec:JCPerturbation}
	Following the notation of \cite{Koch2007}, the Hamiltonian describing a transmon qudit coupled to a readout resonator reads
	\begin{align}
	&H_q+H_r+H_\text{int}= \sum_j\omega_j\ket{j}\bra{j}+\omega_r a^\dagger a\nonumber\\
	&~~~~+\sum_jg_{j,j+1}(a^\dagger\ket{j}\bra{j+1}+a\ket{j+1}\bra{j})\,,\label{eq:QuditResonatorHamiltonian}
	\end{align}
	where $\omega_j$ is the energy (see Appendix~\ref{sec:energylevels}) of the bare qudit state $\ket{j}$, $\omega_r$ is the energy of the readout resonator, and $a^{(\dagger)}$ is its annihilation (creation) operator. The parameters $g_{j,j+1}$ denote generalized Jaynes-Cummings coupling strengths between the qudit and the resonator. Using the approximation $g_{j,j+1}=g\sqrt{j+1}$, the interaction Hamiltonian reduces to $g (a^\dagger b+ab^\dagger)$ \cite{Wallraff2004}, where $b^{(\dagger)}$ is the annihilation (creation) operator of the transmon qudit. The qudit and resonator Hamiltonians $H_q$ and $H_r$ denote two blocks of commuting operators $\ket{i}\bra{j}$ and $a^{(\dagger)}$. The generalized Jaynes-Cummings interaction couples both blocks.
	
	We start with identifying the block-diagonal and block off-diagonal parts,
	\begin{align}
	H_0&=H_q+H_r,~~H_1=H_d\,,\\
	V&=H_\text{int}\,.
	\end{align}
	Using a superposition of all operators appearing in $V$ as an ansatz for $S^{(1)}$,
	\begin{align}
	S^{(1)}&=\sum_j\left(C_j a^\dagger\ket{j}\bra{j+1}-C^*_j a\ket{j+1}\bra{j}\right)\,,
	\label{eq:S1ansatz}
	\end{align}
	the coefficients $C_j$ are obtained as
	\begin{align}
	C_j=\frac{g_{j,j+1}}{\omega_j-\omega_{j+1}+\omega_r}\,.
	\end{align}
	Replacing $C_j$ in Eq.~\eqref{eq:S1ansatz} by this expression and using the definition $g_{-1,0}=0$ as well as the sign convention of \cite{Koch2007,cQED} leads to
	\begin{align}
	\frac12\left[S^{(1)},V\right]&=\sum_{j}\chi_{j-1,j}\ket{j}\bra{j}+\sum_{j}\chi_j a^\dagger a\ket{j}\bra{j}\,.\label{eq:SWTresult}
	\end{align}
	Here,
	\begin{align}
	\chi_{j,j+1}&=\frac{g_{j,j+1}^2}{\omega_{j+1}-\omega_j-\omega_r}\,,\label{eq:chijj1}\\ 
	\chi_j& \equiv\chi_{j-1,j}-\chi_{j,j+1}\nonumber\\ 
	&=\frac{g_{j-1,j}^2}{\omega_{j}-\omega_{j-1}-\omega_r}-\frac{g_{j,j+1}^2}{\omega_{j+1}-\omega_j-\omega_r}\,, \label{eq:chij}
	\end{align}
	and we have neglected terms proportional to $(a^2\ket{j+2}\bra{j}+\mathrm{H.c.})$. This is justified by the possibility to interpret these terms as perturbations that are eliminated by a second Schrieffer-Wolff transformation. This will lead to terms proportional to $\ket{j}\bra{j}$, $a^\dagger a\ket{j}\bra{j}$, $(a^\dagger a)^2\ket{j}\bra{j}$, and also $(a^4\ket{j+4}\bra{j}+\mathrm{H.c.})$. Importantly, for typical values of $g_{j,j+1}$, $\omega_j$, and $\omega_r$, the coefficients of all these terms are a factor of $10^4$ smaller than the previous second-order contributions and can therefore safely be neglected. In conclusion, in Eq.~\eqref{eq:SWTresult}, we arrived at corrections to the Hamiltonian that are diagonal in the qudit and resonator states. The shifts of the qudit and resonator energies are
	\begin{align}
	\tilde{\omega}_j&=\omega_j+\chi_{j-1,j}\,,\nonumber\\ 
	\tilde{\omega}_{r,j}&=\omega_r+\chi_j\,.
	\end{align}
	The resonance frequencies of the qudit transitions ${\ket{i}\leftrightarrow\ket{j}}$ can be estimated to
	\begin{align}
	\tilde{\omega}_{i,j}=\frac{\tilde{\omega}_j-\tilde{\omega}_i}{j-i}\,.\label{eq:omegatilde}
	\end{align}

	\subsection{Qudit Drive}\label{sec:QuditDrivePerturbation}
	In analogy to Appendix~\ref{sec:JCPerturbation}, we perform a Schrieffer-Wolff transformation in a system consisting of a qudit and its drive only. We will use this method to predict the Rabi oscillation frequencies of second-order transitions $\ket{j}\leftrightarrow\ket{j+2}$. For previous work on multiphoton transitions, see, e.g., \cite{MultiPhotonDrive,02Gate}. Starting with the Hamiltonian \cite{cQED},
	\begin{align}
	H&=H_0+\lambda V=\sum_j(\tilde{\omega}_j-\omega_d)\ket{j}\bra{j}\nonumber\\
	&~~~~+\lambda\frac{\Omega_q}{2}\sum_j\sqrt{j+1}(\ee^{-\ii\phi}\ket{j}\bra{j+1}+\ee^{\ii\phi}\ket{j+1}\bra{j})\,,\label{eq:QuditDrive}
	\end{align}
	an expansion in $\lambda$ leads to
	\begin{align}
	H^{(2)}_\text{eff}&=-\frac{\Omega_q^2}{8} \sum_j f_j(\ee^{-2\ii\phi}\ket{j}\bra{j+2} + \ee^{2\ii\phi}\ket{j+2}\bra{j})\,,\label{eq:QuditDrive2ndOrder}
	\end{align}
	where
	\begin{align}
	f_j&=\frac{\sqrt{(j+1)(j+2)}(\tilde{\omega}_{j+2}-2\tilde{\omega}_{j+1}+\tilde{\omega}_j)}{(\tilde{\omega}_{j+2}-\tilde{\omega}_{j+1}-\omega_d)(\tilde{\omega}_{j+1}-\tilde{\omega}_{j}-\omega_d)}\,.
	\end{align}
	Note that these expressions only hold for $\omega_d\neq\tilde{\omega}_{j+1}-\tilde{\omega}_{j}$, i.e., drives that are not resonant with transitions between neighboring qudit levels $\ket{j}\leftrightarrow\ket{j+1}$.

	\begin{figure*}[t]
		\begin{overpic}[width=8cm, tics=20]
			{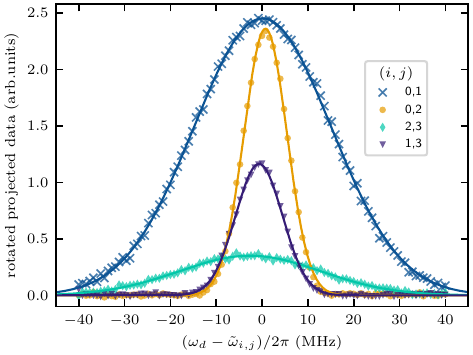}
			\put(25,65){(a)}
		\end{overpic}
		\hfill
		\begin{overpic}[width=8cm, tics=20]
			{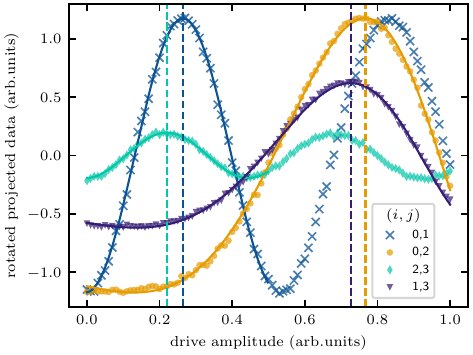}
			\put(25,65){(b)}
		\end{overpic}
		\caption{(a) Qudit resonator spectroscopy of transitions $\ket{i}\leftrightarrow\ket{j}$. Colored markers denote measured data and solid curves correspond to Gaussian fits. We plot each resonance spectrum centered around the predicted transition frequency $\tilde{\omega}_{i,j}$, see Eq.~\eqref{eq:omegatilde}, using $g/2\pi=65\,$MHz. (b) Rabi oscillations $\ket{i}\leftrightarrow\ket{j}$ depending on the drive amplitude for fixed pulse duration. We sweep the readout resonator drive amplitude while keeping all other parameters of the drive fixed. For first-order qudit state transitions, we fit a sinusoidal dependence on a linear function of the drive amplitude in the interval $[0,0.5]$, see the lines connecting crosses and, respectively, diamonds. For second-order qudit state transitions, we fit a sinusoidal dependence on a quadratic function of the drive amplitude in the interval $[0,1]$, see the lines connecting dots and, respectively, triangles. The vertical dashed lines indicate the locations of the first maxima obtained from the fits.}
		\label{fig:calibrations}
	\end{figure*}

	\section{Preparation of Qudit States}\label{sec:StatePreparation}
	We prepare the four lowest Fock states of an IBM Quantum transmon qudit, described in Section~\ref{sec:measurement}, by applying sequences of calibrated $X$ gates to the qudit ground state $\ket{0}$. For simplicity, we implement these $X$ gates via Gaussian pulses. For each pulse, we first calibrate its drive frequency $\omega_d$ and second, its drive amplitude $\Omega_q$.
	
	The optimal drive frequency is obtained from a Gaussian fit to resonance measurement data shown in Fig.~\ref{fig:calibrations}a, where we fix the pulse amplitude to an initial estimate. First, the measured $N=2000$ shots per qudit drive frequency are averaged. Second, these averages are rotated such that their major principal axis is oriented along the $X$ axis. And third, the means are projected onto the $X$ axis which justifies the axis label ``rotated projected data.'' For the spectroscopy measurements, in addition, we define the origin of the $y$ axis of Fig.~\ref{fig:calibrations}a to correspond to the initial state of the analyzed transition and the maximum to the final state. Our estimated frequency $\tilde{\omega}_{i,j}$ is calculated by Eq.~\eqref{eq:omegatilde} using $g/2\pi=65\,$MHz.
	
	To obtain the initial estimate of the qudit drive amplitude, we define the rotation angle $\theta$ of a resonant Rabi oscillation between states $\ket{j}$ and $\ket{k}$. Comparing both sides of
	\begin{align}
	\exp(-\ii H t)&=\exp\left(-\frac{\ii}{2} \theta (\ket{j}\bra{k}+\ket{k}\bra{j})\right)\,,
	\end{align}
	where the left-hand side is the time evolution of the effective drive Hamiltonian,
	\begin{align}
	H&=\Lambda(j,k)(\ket{j}\bra{k}+\ket{k}\bra{j})\,.
	\end{align}
	Therefore, the rotation angle $\theta$ depends on the effective Rabi frequency $\Lambda(j,k)$ and the pulse duration $t$. Following Eq.~\eqref{eq:QuditDrive} for $\lambda=1$, the rotation angle $\theta$ for Rabi oscillations between $\ket{j}$ and $\ket{j+1}$ is given by
	\begin{align}
	\theta&=t_{j,j+1}\Omega^{(j,j+1)}\sqrt{j+1}\,.\label{eq:theta1}
	\end{align}
	The rotation angle for Rabi oscillations between non-neighboring states $\ket{j}$ and $\ket{j+2}$ can be computed using Eq.~\eqref{eq:QuditDrive2ndOrder},
	\begin{align}
	\theta&=t_{j,j+2}\frac{\left(\Omega^{(j,j+2)}\right)^2}{4}f_j\,.\label{eq:theta2}
	\end{align}
	The Rabi frequency of the $\ket{0}\leftrightarrow\ket{2}$ transition scales quadratically with $\Omega^{(0,2)}$. Using Eqs.~\eqref{eq:theta1} and \eqref{eq:theta2}, the initial estimate for the $\pi$-pulse amplitudes $\Omega^{(j,k)}_\pi$ can be related to the default $X$-gate amplitude $\Omega^{(0,1)}_\pi$ reported by the IBM Quantum backend,
	\begin{align}
	\Omega^{(j,j+1)}_\pi&=\frac{\Omega^{(0,1)}_\pi}{\sqrt{j+1}}\,,\label{eq:Om01}\\
	\Omega^{(j,j+2)}_\pi&=2\sqrt{\frac{\Omega^{(0,1)}_\pi t_{0,1}}{f_jt_{j,j+2}}}\,.\label{eq:Om02}
	\end{align}
	Here we used that all single-qudit operations are implemented within the same duration $t_{j,j+1}=t_{0,1}$.
	
	After evaluating the resonance measurement, we calibrate the $X$-gate drive amplitude via Rabi oscillations, see Fig.~\ref{fig:calibrations}b. The data are rotated and projected onto the major principal axis as described before for the spectroscopy measurements. As shown in Eq.~\eqref{eq:QuditDrive2ndOrder}, the Rabi frequency for transitions $\ket{j}\leftrightarrow\ket{j+2}$ depends nonlinearly on the drive amplitude. Since these transitions are suppressed by the small factor $\Omega_q f_j$, we choose $t_{j,j+2}=2t_{0,1}$ such that $\Omega^{(0,2)}_\pi$ does not exceed the limits of IBM Quantum software/hardware restrictions. The $\pi$ amplitude of an $X$-gate pulse is identified with the location of the first maximum in Fig.~\ref{fig:calibrations}b, indicated by dashed lines. For transitions between neighboring states, we fit a sine dependence on a linear function of $\Omega_q$ and, for second-order transitions, we fit a sine dependence on a second-order polynomial of $\Omega_q$. Using those fits, any desired rotation angle, e.g., $\pi$ for an $X$ gate or $\pi/2$ for a Hadamard gate, can be mapped back to a corresponding pulse amplitude.
	
	The sequence of calibrating drive frequency and amplitude described above can be iterated several times to improve gate fidelity. For simplicity, we consider only one round of calibrations. To increase fidelity, we chose the initial value for $\Omega_{1,3}$ based on prior test measurements.
	
	Implementing gates in the $\ket{j}\leftrightarrow\ket{j+2}$ subspace results in two advantages. First, our implementation of an $X$-gate $X_{j,j+2}$ between $\ket{j}$ and $\ket{j+2}$ takes only twice the single-qudit gate duration $t_{0,1}$. In contrast, using single-qudit gates, $X_{j,j+2}$ consists of three single-qudit operations $X_{j,j+1}X_{j+1,j+2}X_{j,j+1}$ with a total duration of $3t_{0,1}$. Second, the calibration of the drive frequencies (amplitudes) for $\ket{0}\leftrightarrow\ket{1}$ and $\ket{0}\leftrightarrow\ket{2}$ are independent of each other and can therefore be combined into a single \textsc{Qiskit} job (set of measurements submitted to an IBM Quantum device). In contrast, the frequency calibration for the transition $\ket{1}\leftrightarrow\ket{2}$ depends on the Rabi measurement for the transition $\ket{0}\leftrightarrow\ket{1}$. In total, we can perform our calibration in four \textsc{Qiskit} jobs: (\textit{i}) drive frequency of $X_{01}$ and $X_{02}$, (\textit{ii}) drive amplitude of $X_{01}$ and $X_{02}$, (\textit{iii}) drive frequency of $X_{12}$, $X_{23}$, and $X_{13}$, and (\textit{iv}) drive amplitude of $X_{12}$, $X_{23}$, and $X_{13}$. In contrast, the standard sequential calibration of single-qudit $X$ gates would take six jobs: two for each of the three single-qudit $X$ gates between neighboring states.

	\section{Dirichlet Distribution}\label{sec:DirichletDistribution}
	We describe the probability distribution that estimates the populations $p_j\equiv|c_j|^2$ of the state we want to analyze using Bayesian inference. To this end, we define a recursion relation
	\begin{align}
	P^{(j+1)}(\vec{p})&=\frac{P(z_j|\vec{p})}{P(z_j)}P^{(j)}(\vec{p})\,,\\
	P(z_j)&=\int\d^{d-1}p\,P(z_j|\vec{p})P^{(j)}(\vec{p})\,,
	\end{align}
	between the estimated probability distribution (prior) $P^{(j)}(\vec{p})$ of the Gaussian amplitudes $\vec{p}$ before and after receiving the $j^\text{th}$ data point $z_j$, also called shot. Each shot follows the probability distribution $P(z_j|\vec{p})$ defined in Eq.~\eqref{eq:ProbToLearn}. After obtaining $N$ data points, this leads to the probability distribution $P^{(N)}(\vec{p})$,
	\begin{align}
	P^{(N)}(\vec{p})&=\frac{\prod_{j=0}^{N-1}P(z_j|\vec{p})}{\prod_{j=0}^{N-1}P(z_j)}P^{(0)}(\vec{p})\,.\label{eq:probN1}
	\end{align}
	The initial prior $P^{(0)}(\vec{p})$ is chosen to be a uniform distribution.
	
	If the width of the Gaussians is small compared to their distances, this method is equivalent to the description given in the following, see, also, Section~\ref{sec:FiniteSampling}. For simplicity, instead of using this Bayesian ansatz of the probability distribution, each of the $N$ shots is classified as one of the $d$ qudit states. The components $N_j$ of $\vec{N}$ equal the number of shots classified as $\ket{j}$. This procedure corresponds to neglecting the position $z_j$ of this shot, i.e., the Gaussian weight $G(z,A_j,\sigma_j)$ of the shot. Given $N_j$, the posterior probability distribution $P^{(N)}(\vec{p})$ for the qudit populations $p_j$ for a perfect measurement is equal to
	\begin{align}
	P(\vec{p} \,| \vec{N}) &= \mathrm{Dir}(\vec{p},\vec{N}) = \frac{(N+d-1)!}{\prod_{k=0}^{d-1}N_k!} \prod_{k=0}^{d-1} p_k^{N_k}\,,
	\end{align}
	with
	\begin{align}
	\sum_{j=0}^{d-1}p_j &=1,~~\sum_{j=0}^{d-1}N_j=N,
	\end{align}
	introducing the Dirichlet distribution $\mathrm{Dir}$ \cite{DirichletDist}. The location of its maximum (also called mode) with respect to $p_j$ is given by $N_j/N$, and its variances follow as
	\begin{align}
	\mathrm{Var}[p_j] = \frac{(N_j+1)/(N+d)[1-(N_j+1)/(N+d)]}{N+d+1}\,.
	\end{align}
	Defining $n_j=N_j/N$, for large $N$ the variance of $p_j$ scales like $n_j(1-n_j)/N$.
	
	We now consider the assignment matrix $M$, see Eq.~\eqref{eq:AssignmentMatrix}, which describes misclassification errors. Using a Bayesian posterior ansatz, we find that the probability distribution has to be modified to
	\begin{align}
	P(\vec{p}\,|\vec{N}) &= \frac{1}{\mathcal{N}} \mathrm{Dir}(M \vec{p}, \vec{N})\label{eq:DirwithM}
	\end{align}
	where
	\begin{align}
	\mathcal{N}&=\int\limits_{V_{\vec{p}}}\d^dp\,\mathrm{Dir}(M \vec{p}, \vec{N})\,.
	\end{align}
	The assignment matrix maps proper states $\vec{p}$ from $V_{\vec{p}}$ (related to a $(d-1)$-simplex) to a subspace $V_{M\vec{p}}\subseteq V_{\vec{p}}$. Therefore, if $\vec{N}/N\in V_{M\vec{p}}$, the mode of Eq.~\eqref{eq:DirwithM} is naturally $M^{-1}\vec{N}/N$.

	\section{Analytical Assignment Matrix}\label{sec:OwenT}
	In this appendix, we present an analytical expression for the assignment matrix in specific setups that allows for a fast numerical implementation. The following expressions hold for systems with $\sigma_j=\sigma$. Furthermore, all states have to lie on a circle centered at $A_c=x_c+\ii y_c$. Examples are qudit systems with $\omega_m=\omega_d$, cf.~discussion in Section~\ref{sec:rotframedrive}, or qutrit systems with arbitrary $\omega_m$. Equation~\eqref{eq:AssignmentMatrix} can be written as 
	\begin{align}
	M_{i,j}&=\frac14\left(1-\mathrm{erf}\left(\frac{x_j-x_c}{\sqrt{2}\sigma}\right)\right)\nonumber\\
	&~~+T\left(\frac{x_j-x_c}{\sigma}, -a_i, \frac{y_c-y_j+a_i(x_j-x_c)}{\sigma}\right)\,,
	\end{align}
	where the slope of the bisecting line between $A_i$ and $A_{i-1}$ is
	\begin{align}
	a_i&=-\frac{x_{i}-x_{i-1}}{y_{i}-y_{i-1}}\,.
	\end{align}
	Here, the variables $x_j$ and $y_j$ are the real and imaginary part of $A_j$. The coordinate system is rotated such that $A_i$ and $A_{i+1}$ are aligned along the real axis in phase space.
	\onecolumngrid
	Finally,
	\begin{align}
	&T(h,a,b)=\frac{1}{2\sqrt{2\pi}}\int\limits_h^\infty\d x\,\exp\left(-\frac{x^2}{2}\right)\mathrm{erf}\left(\frac{a x+b}{\sqrt{2}}\right)=\frac14\mathrm{erf}\left(\frac{b}{\sqrt{2(1+a^2)}}\right)\left(1-\mathrm{erf}\left(\frac{h}{\sqrt{2}}\right)\right)\nonumber\\
	&~~+T\left(\frac{b}{\sqrt{1+a^2}},a+\frac{h(1+a^2)}{b}\right)+T\left(h,a+\frac{b}{h}\right) -T\left(\frac{b}{\sqrt{1+a^2}},\frac{h\sqrt{1+a^2}}{b}\right)-T\left(h,\frac{b}{h\sqrt{1+a^2}}\right)\,,
	\end{align}
	is a generalized version of Owen's $T$ function, $T(h,a)=T(h,a,0)$ \cite{OwenT,OwenTgeneral}.
	\twocolumngrid

\end{document}